\documentclass[conference]{IEEEtran}
\IEEEoverridecommandlockouts

\usepackage{graphicx} 
\usepackage{amsmath} 
\usepackage{amssymb}  
\usepackage{comment} 
\usepackage{hyperref}
\usepackage{multirow}
\usepackage{stfloats}
\usepackage{algorithm}
\usepackage{algpseudocode}
\usepackage{tabularx}
\usepackage[table]{xcolor}
\usepackage{xcolor}
\usepackage[giveninits=true,isbn=false,style=ieee]{biblatex}

\bibliography{ref}
\AtBeginBibliography{\footnotesize}

\DeclareMathOperator*{\argmax}{arg\,max} 
\DeclareMathOperator*{\argmin}{arg\,min}

\algrenewcommand\algorithmicrequire{\textbf{Input:}}
\algrenewcommand\algorithmicensure{\textbf{Output:}}

\makeatletter
\newcommand{\multiline}[1]{\begin{tabularx}{\dimexpr\linewidth-\ALG@thistlm}[t]{@{}X@{}}#1\end{tabularx}}
\makeatother

\title{Proximal Policy Optimization Based Reinforcement Learning for Joint Bidding in Energy and Frequency Regulation Markets%
\thanks{*Corresponding author: Hao Wang.}}

\author{%
\IEEEauthorblockN{Muhammad Anwar}
\IEEEauthorblockA{%
\textit{Faculty of Information Technology} \\
\textit{Monash University}\\
Melbourne, Australia \\
\texttt{manw0002@student.monash.edu}}
\and
\IEEEauthorblockN{Changlong Wang}
\IEEEauthorblockA{%
\textit{Department of Civil Engineering} \\
\textit{Monash University}\\
Melbourne, Australia \\
\texttt{chang.wang@monash.edu}}
\and
\IEEEauthorblockN{Frits de Nijs, Hao Wang*}
\IEEEauthorblockA{%
\textit{Department of Data Science and AI} \\
\textit{Monash University}\\
Melbourne, Australia \\
\texttt{\{frits.nijs,hao.wang2\}@monash.edu}}
}

\begin{document}
\maketitle

\begin{abstract}
Driven by the global decarbonization effort, the rapid integration of renewable energy into the conventional electricity grid presents new challenges and opportunities for the battery energy storage system (BESS) participating in the energy market. Energy arbitrage can be a significant source of revenue for the BESS due to the increasing price volatility in the spot market caused by the mismatch between renewable generation and electricity demand. In addition, the Frequency Control Ancillary Services (FCAS) markets established to stabilize the grid can offer higher returns for the BESS due to their capability to respond within milliseconds. Therefore, it is crucial for the BESS to carefully decide how much capacity to assign to each market to maximize the total profit under uncertain market conditions. This paper formulates the bidding problem of the BESS as a Markov Decision Process, which enables the BESS to participate in both the spot market and the FCAS market to maximize profit. Then, Proximal Policy Optimization, a model-free deep reinforcement learning algorithm, is employed to learn the optimal bidding strategy from the dynamic environment of the energy market under a continuous bidding scale. The proposed model is trained and validated using real-world historical data of the Australian National Electricity Market. The results demonstrate that our developed joint bidding strategy in both markets is significantly profitable compared to individual markets.
\end{abstract}

\section{Introduction} \label{intro}

\subsection{Background}
The rapid integration of renewable energy into the conventional electricity grid poses a unique challenge to energy market operators. The variability of solar and wind energy often triggers a mismatch between renewable generation and consumption, which results in price volatility in real-time energy markets \cite{wang_review_2015}. The battery energy storage system (BESS) can take advantage of this volatility to make a profit through arbitrage \cite{bradbury_economic_2014}, by charging the battery at a low price and discharging at a higher price to retain a profit margin. 

With the increasing penetration of renewable generation from solar photovoltaic and wind turbine generators in modern power grids, the task of balancing electricity demand and supply in real-time has become increasingly challenging. To ensure system reliability, it is critical to balance distributed energy resources and utility-scale generation with electricity consumption across the entire grid at all times. The ancillary energy market provides frequency regulation services to ensure power system security in the face of limited dispatchability of the renewable supply. BESS is especially incentivized to participate in the frequency regulation market due to its fast response capability and the associated high economic return~\cite{FCAS_Model_in_NEMDE}. 

Leveraging its flexibility, BESS can charge and discharge rapidly in both markets in response to price and control signals. However, the market environments are stochastic, making BESS scheduling a very challenging problem. As such, our paper aims to develop a deep reinforcement learning (DRL)-based bidding strategy to unlock the value of BESS in both markets. This could help mitigate capital investment risk and contribute to market stability and efficiency.

\subsection{Related Work}
Several optimization-based methods in literature address the optimal energy market bidding problem by employing bi-level optimization, with the mathematical program with equilibrium constraint dominating the literature as the most explored approach \cite{nemati2018optimization, kardakos_optimal_2014}. These methods are computationally highly intensive with no guaranteed upper bound of convergence to the optimal solution. While various evolutionary approaches explored in the literature (\cite{shah2019optimal, reddy2017bidding}) can converge faster, such offline algorithms cannot cope with the sudden change of the price forecasts in a highly volatile energy market. Other works adopt a game-theoretic approach to model the intricate relationship among the market participants, in order to compute price at market equilibrium \cite{abapour2020game, wang_incentivizing_2018}. However, the resulting bi-level optimization problems are challenging to solve and require detailed financial knowledge of the rival participant's bids to be available to each other, which is not usually available in a realistic market. 

More recent studies have used reinforcement learning to develop online bidding strategies. A number of works have addressed energy market optimization using different approaches focusing separately either on energy arbitrage \cite{wang_energy_2018,xu_arbitrage_2019,pedasingu_bidding_2020} or frequency regulation market \cite{dong_strategic_2021,chen_bargaining_2021}. Most related works attempted to optimize the bidding strategy for a discrete range of market prices that can provide the buy and sell signal over a price range but cannot decide the optimal price point. This research uses a function approximation technique to overcome the limitation of finite action space by integrating Deep Neural Networks (DNN).  

As discussed, a vast majority of the existing literature addressed the BESS bidding problem under heroic assumptions, with some assuming participants have perfect foresight of the market prices. Our proposed DRL-based bidding strategy is more suitable for the BESS in a complex dynamic market environment since the algorithm is online by nature and can adapt well to a frequent unexpected shift in the underlying distribution of the market state. Furthermore, most existing studies focus on either a single market using reinforcement learning or multiple markets using optimization methods based on market prices, leaving a research gap of how to bid in multiple markets in real-time optimally. Our work bridges the gap by developing a proximal policy optimization-based reinforcement learning algorithm, enabling the BESS to bid concurrently in energy and frequency markets.

\subsection{Main Results and Contributions}
This paper proposes a Markov Decision Process (MDP) based approach capable of making optimal bidding decisions in the highly dynamic environment of the energy market. A reinforcement learning (RL) model is formulated to optimize the bidding strategy of the BESS in the real-time energy and FCAS markets. The proposed reinforcement-learning algorithm based on proximal policy optimization (PPO) jointly optimizes the BESS bidding strategy in energy and FCAS markets to exploit the maximum flexibility among both markets. DNN is used to extend the RL model to the continuous action-state space of the energy market. The proposed framework is validated with real-world data of the Australian National Electricity Market, which shows that the joint bidding strategy yields higher profit than the individual markets combined.

The rest of the paper is organized as follows. Section~\ref{sec:formulation} provides an overview of the Australian energy market and presents the system model. The PPO-based reinforcement learning algorithm is presented in Section~\ref{sec:RL}. Experimental results and analysis are discussed in Section~\ref{sec:results}. Section~\ref{sec:conclusion} concludes our work with future research directions.

\section{Problem Formulation}\label{sec:formulation}
\subsection{National Electricity Market}
The National Electricity Market (NEM) \cite{noauthor_nem_nodate} in Australia allows electricity trading between generators and retailers through an interconnected grid of five jurisdictions on the east coast of Australia. 
The spot market is managed by the Australian Energy Market Operator (AEMO) to ensure power supply and demand is matched simultaneously. 

In the spot market, power generators submit an `offer' indicating supply of a specified amount of electricity for a specified period of time at a specified `bid' price. To meet demand in a least-cost manner, AEMO prioritizes the offers of the lowest bidding generators. National Electricity Market Dispatch Engine (NEMDE), as the central dispatch process of the NEM, calculates the `spot' price every five minutes \cite{noauthor_5min_nodate} by optimizing electricity production to actual demand subject to a large number of generation and transmission constraints. 

The Frequency Control Ancillary Services (FCAS) market is dedicated to maintaining the system frequency within $49.85$ to $50.15$ Hz \cite{FCAS_Guide_AEMO}. 
In the FCAS market, participants submit a joint FCAS offer comprising ten price bands, with each specifying the FCAS availability limit and the corresponding price per unit of energy generation capacity. NEMDE jointly optimizes the overall energy consumption and the FCAS requirements subject to technical limitations of these services at the lowest possible cost by dispatching the lowest offers from the energy and FCAS markets available to meet the combined energy requirements. This is the market context that our model is particularly designed for.

\subsection{Joint Profit Maximization Problem}
In this section, we introduce model formulations for BESS profit maximization in a joint market, where the BESS can participate in both energy arbitrage and frequency regulation services. 
Throughout this paper, we use the subscript $t$ to denote a variable in time slot $t$ corresponding to the 5-minute settlement period for both energy and FCAS regulation markets.

\subsubsection{Energy Market}
The BESS participates in the energy arbitrage, intending to maximize its arbitrage profit by charging the battery at low prices and discharging at higher prices in the energy market \cite{wang_energy_2018}. In each time slot $t$, the BESS needs to decide how much energy $C_t^e$ to charge into or how much energy $D_t^e$ to discharge from the energy storage depending on the market price constituting the market offers for that time slot. However, it can not charge and discharge energy simultaneously from the single shared energy storage, ensured by the non-negative binary variables $b_t^c = \{ 0,1\}$ and $b_t^d = \{ 0,1\}$. Let $\eta_c \in (0,1)$ and $\eta_d \in (0,1)$ denote the charging and discharging efficiency of the energy storage, respectively, and $\rho_t^e$ denote the clearing price of the wholesale energy market, the energy arbitrage revenue generated from the energy market is
\begin{equation}
        R^e = \sum_{t=1}^T \rho_t^e\left(b_t^d \eta_d D^e_t -  b_t^c \frac{1}{\eta_c} C^e_t \right),
\end{equation}
where $\eta_d D^e_t$ and $\frac{1}{\eta_c} C^e_t$ denote discharge and charge energy measured from the outside of the BESS.

The BESS charge and discharge should satisfy the following constraint to avoid charging and discharging simultaneously in any time slot $t$ due to sharing one single energy storage, i.e.,
\begin{equation}
       b_t^c +  b_t^d \leq 1,
\end{equation}

\subsubsection{FCAS Market}
While participating in the FCAS market, the BESS will receive normalized regulation signals $S_t = {S_t \in [-1, 1]}$ from the market operator, which indicates the enabled amount of frequency regulation. For example, the positive value indicates the Raise Regulation signal, and the negative value indicates the Lower Regulation signal. For each time slot $t$, the BESS decides the capacity $R_t^f$ to bid for the Raise regulation and the capacity $L_t^f$ for the Lower regulation. We denote the FCAS prices for raise regulation and lower regulation at time $t$ are $\rho_t^r$ and $\rho_t^l$, respectively. Therefore, the revenue of bidding for the raise and lower regulation services in the FCAS market is
\begin{equation}
R^{\text{FCAS}} = \sum_{t=1}^T{\left( \rho_t^r R_{t}^f + \rho_t^l L_{t}^f\right)}.
\end{equation}

Given the FCAS bidding, in each time slot $t$, the BESS need to discharge (e.g., $S_t R_t^f$) or charge (e.g., $ - S_t L_t^f$), following the direction of the normalized regulation signals $S_t$ scaled proportionally to the bid capacity in order to generate the frequency response \cite{chen_bargaining_2021}.

\subsubsection{The BESS Model}
The BESS continues to participate in the two markets simultaneously, the energy stored in the BESS, denoted by $E_t$, also changes, however, within limits, i.e., \(0 \leq E_{\min} \leq E_t \leq E_{\max}\), where $E_{\min}$ and $E_{\max}$ are the lower bound and upper bound for $E_t$. The lower bound and upper bound are often set as $10\%$ and $90\%$ of the BESS capacity. The operational constraints for the BESS are as follows.
\begin{align}
    & E_{t+1} = E_t + b_t^c(C_t^e - S_t L_t^f) - b_t^d (D_t^e + S_t R_t^f), \\
    & E_{\min} \leq E_t \leq E_{\max},
\end{align}
where variables $b_t^c$ and $b_t^d$ are used to avoid charging and discharging at the same time across FCAS and energy arbitrage.

The actual charging and discharging of energy during trading results in permanent damage to the BESS. The BESS degradation cost $C^{\text{BESS}}$, is modeled with the total discharged energy for both the energy and FCAS markets using a battery technology specific cost-coefficient $\alpha^{\text{BESS}}$ as
\begin{equation}
C^{\text{BESS}} = \alpha^{\text{BESS}} \sum_{t=1}^T{ \left( D_{t}^e + S_t R_t^f \right)}.
\end{equation}

The objective of the BESS is to maximize the revenue of participating in the wholesale and FCAS markets over an operational horizon up to $T$, i.e., 
\begin{equation}
\max~R^e + R^{\text{FCAS}} - C^{\text{BESS}}.
\label{eq:optObj}
\end{equation}

This optimal bidding problem formulated above is a mixed-integer linear programming (MILP) problem, which can be solved for the optimal charge or discharge decisions corresponding to the optimal amount of energy to bid for the specific energy markets at the current market-clearing price. However, an offline solution requires the energy market price to be known or a reasonable price forecast to be available, which is extremely difficult with the energy market dynamics. Therefore, an online method should be adopted to solve the bidding optimization problem efficiently, which motivates us to develop a reinforcement learning algorithm in Section~\ref{sec:RL}.

\section{Reinforcement Learning Algorithm}\label{sec:RL}
This paper proposes reinforcement learning (RL) as the solution method for the optimal bidding problem and Proximal Policy Optimization (PPO) as the RL algorithm of choice. In this section, the MDP of the bidding problem is described along with the proposed RL method.

\subsection{Markov Decision Process Formulation}
An MDP formally describes the environment for reinforcement learning. At every time step, the process is observed in one of several possible states $s_t$, based upon which a decision of choosing an action $a_t$, is to be made, resulting in transitioning to a different state $s_{t+1}$ and whereby receiving a corresponding reward $r_t(s_t, a_t)$. The MDP is described as a tuple of five, $(S, A, P, R, \gamma)$ \cite{sutton_reinforcement_2018}, where $S$ is a finite set of all possible states attainable within the environment, $A$ is a finite set of actions that an agent can take to move between the states, $P$ is the state transition probability matrix where each entry containing $P(s_{t+1} | s_t, a_t)$, the probability of taking action resulting in a particular state transition, $R$ is the reward function awarded to the agent upon taking action resulting in state s and $\gamma \in [0, 1] $ is the discount factor. 

\subsubsection{State Space}
The state of the MDP at time instance $t$ is defined $s_t = (E_t, \rho_t^e, \rho_t^r, \rho_t^l)$. Each state of the process can be fully described by the energy stored in the battery, energy price of the spot market, the price of the Regulation Raise market, and the price of the Regulation Lower market, respectively. The designed state meets the Markovian property so that $s_t$ depends only on $s_{t-1}$ and is independent of all previous states. 

\subsubsection{Action Space}
Depending on the current state $s_t$, the agent takes action $a_t = (a_t^f, a_t^e)$. Note that $a_t^f \in [-E_{t-1}, (E_{\max} -E_{t-1})]$ represents the amount of energy to bid in the FCAS market, where a positive action indicates availability for the Regulation Lower market, negative action indicates availability for the Regulation Raise market, and zero indicates the BESS is not participating in the FCAS market\footnote{We assume that the BESS has a high C rate, such that the BESS can charge or discharge fully in a single time slot. We will consider the general setting with an arbitrary C rate in our future work.}. Then, $a_t^e \in [-(E_{t-1} -  |a_t^f|), (E_{\max} - |a_t^f|))] $ represents the amount of remaining energy to bid for the energy market, where a positive action indicates charge, negative action indicates discharge, and zero indicates neither charge nor discharge. Importantly, the agent must ensure $a_t^f \cdot a_t^e \geq 0 $ to avoid generating simultaneous charge and discharge action pairs.

\subsubsection{Reward}
Upon taking the action $a_t$ at the state $s_t$ leading to the state transition $s_{t+1}$, the agent receives the reward $r_t$, which can be defined by the revenue generated at time $t$ according to (\ref{eq:optObj}) as
\begin{equation}
\begin{aligned}
r_t &= {\rho_t^e\left(\eta_d D^e_t -  \frac{1}{\eta_c} C^e_t\right)} +
 \rho_t^r R_{t}^f + \rho_t^l L_{t}^f \\
 & -  \alpha^{\text{BESS}} \left(D_{t}^e + S_t R_t^f \right).
\label{eq:reward}
\end{aligned}
\end{equation}

The objective of the BESS is to obtain the optimal bidding policy $\pi(s_t, \theta)$ that maximizes the reward $r_t$, which is solved using the RL algorithm as described in the following section.



\subsection{Solution Method}
The Proximal Policy Optimization (PPO) \cite{schulman_proximal_2017} is a policy gradient-based reinforcement learning algorithm. Earlier policy gradient-based DRL algorithms are difficult to implement \cite{mnih_asynchronous_2016}. The newer Trust Region Policy optimization \cite{schulman_trust_2017} requires higher-order Kullback–Leibler (KL) constraints, which makes training resource intensive. The PPO algorithm addresses this limitation using a clipped surrogate objective. 

\begin{figure}[!bhp]
    \centering
    \vspace{-3mm}
    \includegraphics[width=\columnwidth]{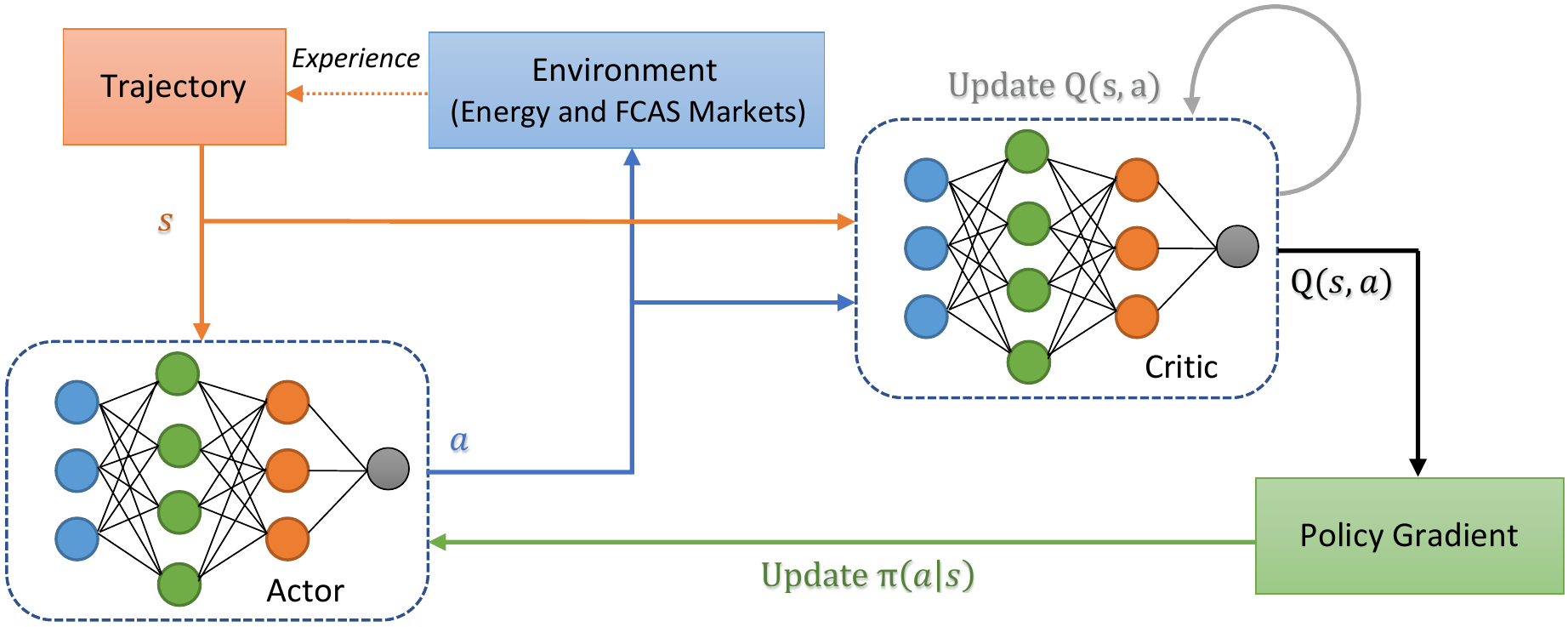}
    \vspace{-5mm}
    \caption{High-level diagram of the proximal policy optimization algorithm.}
    \label{fig:PPODiagram}
\end{figure}

The PPO falls into the Actor-Critic family of RL algorithms comprising of two neural networks as shown in Fig.~\ref{fig:PPODiagram}. The actor is the policy network that determines the policy function $\pi_\theta(s, a)$, and the critic evaluates the selected policy using the state-value function estimation $\hat{V}_\phi^\pi(s)$.  

The model parameters of the actor and critic networks, respectively $\theta$ and $\phi$, are trained on the energy market data using the PPO algorithm. For each iteration $k$ of training, instead of exploring the full episode, $D$ finite length trajectories are sampled from the training data using the current policy $\pi_\theta$. The simulation generates $T$ bidding decisions $a_t$ for the energy and frequency regulation markets based on the current policy for each trajectory. This results in profit or loss of $r_t$ for each time-step $t$ depending on the state $s_t$ using the reward function defined in (\ref{eq:reward}). The expected discounted reward of the trajectory is estimated by the value function 
\[{V}_\phi^\pi(s_{t}) = \mathbb{E}\biggl[\sum_{k=0}^T \gamma^k r_{t+k} ( s_{t+k}, a_{t+k})\biggr].\]

The profitability of the bidding strategy $\pi$ is computed by the advantage function, which is used by the policy network to update the gradient of the policy parameters towards the optimal policy.  The generalized advantage estimation technique \cite{mnih_asynchronous_2016} is used to estimate advantage function $\hat{A}_t$ for the policy truncated to $T$ time-steps,
\[\hat{A}_t = \delta_t + (\gamma \lambda) \delta_{t + 1} + \ldots + (\gamma \lambda)^{T-t+1} \delta_{T-1},\]
where $\delta_t = r_t + \gamma V(s_{t+1}) - V(s_t)$, and $\lambda \in [0, 1]$ is a weighting parameter. Since actor and critic networks are independent, their parameters $\theta$ and $\phi$ are optimized separately using mini-batch gradient descent. Specifically, $\phi$ is updated using a value loss function with the objective,
\[\phi_{k+1} = \argmin_{\phi} \frac{1}{|D|T} \sum_D \sum_{t=0}^T {\left(V_{\phi_k}^\pi(s) - \hat{V^\pi}(s)\right)}^2.\]

In PPO, the updates to the policy network are clipped using a hyper-parameter $\epsilon$, so that the probability ratio ${r(\theta)} = \frac{\pi_\theta(a | s)}{\pi_{\theta_{old}}(a | s)}$ is constrained within the interval $(1-\epsilon, 1 + \epsilon)$. This prevents potentially large policy updates and provides better stability during training. Hence, $\theta$ is updated as
\begin{equation}
\begin{aligned}
	\theta_{k+1} =& \argmax_{\theta} \frac{1}{|D|T} \cdot {}\\&\sum_D \sum_{t=0}^T { 
    \min\left(
        {r(\theta)} {\hat{A_t}},
        \text{clip}({r(\theta)},1\!-\!\epsilon,1\!+\!\epsilon) {\hat{A_t}}
    \right).}
\end{aligned}
\end{equation}


\section{Experiments and Results} \label{sec:results}
In this section, we evaluate the performance of the proposed RL-based bidding algorithm in the context of real-world market data collected from the Australian National Electricity Market \cite{noauthor_nem_market_data_nodate}.

\subsection{Simulation Setup}
For the experiments, the historical NEM market data published by AEMO of the last five years (2015-2020) are used \cite{noauthor_nem_market_data_nodate}. The data set is collected in 5-minute intervals to be consistent with the current market settlement period \cite{noauthor_5min_nodate}. For the simulation, a BESS with the capacity of $100$MWh, charging efficiency $\eta_c = 0.95 $, discharging efficiency $\eta_d =0.95$, and battery specific cost-coefficient $\alpha^{\text{BESS}} =\$1/$MW is considered. As the NEM operates separate markets in five different regions, the energy market of Victoria is considered for the simulation unless otherwise specified. The market closing prices from the wholesale energy market and the Regulation Raise and Regulation Lower FCAS markets at every 5-minute interval are collected. In addition to real market price, synthetic FCAS regulation signal is used for the purpose of simulation, which is generated adding random noise ${\displaystyle X \sim {\mathcal {N}}(0 ,0.5 ^{2})}$ within  contingency frequency band (e.g., 49.5Hz to 50.5Hz). Given the bidding in the FCAS market, the BESS follows the regulation signal to charge and discharge, and jointly optimizes the schedule in the energy market.

The actor and critic networks are constructed with an input layer followed by two hidden layers, each of $512$ in length, and finally an output layer. The width of the actor and critic networks are determined by the number of states and actions, respectively. Input and hidden layers are activated with Rectified Linear Unit (ReLU), while the output layer uses a $\mathit{tanh}$ activation function. Adam optimizer is used for training the network using stochastic gradient descent with batch size $256$ and a smooth $L1$ loss function. The PPO algorithm is implemented using Python, PyTorch, and ElegantRL~\cite{erl}. Market data from the entire year of 2016 is used to train the model. The data from the first quarter of 2017 is used to evaluate the performance measured by the cumulative profit gained within the period.

\subsection{Results and Analysis}
The proposed reinforcement learning-based framework is evaluated in the energy and FCAS markets, and the results are presented as follows.

\begin{figure}[h]
    \centering
    \vspace{-3mm}
    \includegraphics[width=0.46\textwidth]{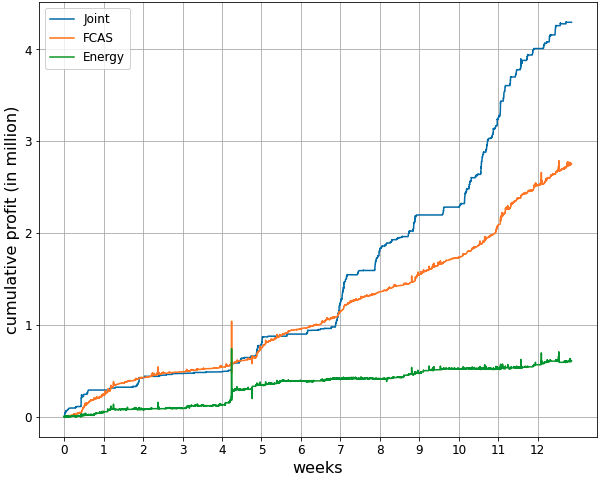}
    \vspace{-2mm}
    \caption{Comparison of cumulative return gained from energy, FCAS and joint markets.}
    \label{fig:jointReturn}
\end{figure}

The proposed RL-based bidding framework is used to participate simultaneously in the energy market and the two FCAS (e.g., Regulation Raise and Regulation Lower) markets. The single BESS is shared to jointly bid in both markets, which yields the maximum profit, compared to the profits earned from bidding individually in the energy or FCAS markets (See Figure~\ref{fig:jointReturn}). Surprisingly, the long-term profit generated from the joint bidding can be higher than the sum of the profits generated from participating individually in the two markets. This demonstrates that the proposed PPO RL-based bidding algorithm can effectively discover and exploit the flexibility in simultaneous bidding in both markets, further improving the viability of energy storage.

\section{Conclusion}\label{sec:conclusion}
In this paper, an MDP was formulated to examine the optimal bidding decisions of BESS participating in the energy and regulation markets. We developed a deep reinforcement learning method jointly optimizing the BESS bidding decisions using proximal policy optimization. Our proposed method was evaluated based on historical market data from the Australian National Electricity Market. The results demonstrated that the proposed method, which jointly optimizes the profit in both the energy and the FCAS market, can yield higher profitability than in individual markets combined.

Future work will incorporate a non-linear BESS degradation model with real-world regulation signals to further evaluate regulation services in different time scales.

\printbibliography

\end{document}